\documentclass[a4paper,12pt]{article}
\usepackage{graphicx}
\usepackage[body={6.9in, 9.8in},left=1.0in, top = 1.0in]{geometry}
\usepackage{epsfig}
\usepackage{appendix}
\usepackage{subfigure}
\begin{document}
\title{ Form factors and charge radii in a QCD inspired potential model using the  Variationally Improved Perturbation Theory (VIPT) }
\author{$^{1}$Bhaskar Jyoti Hazarika and $^{2}$D K Choudhury \\
$^{1}$Dept of Physics,Pandu College,Guwahati-781012,India\\
e-mail:bh53033@gmail.com\\
$^{2}$Dept. of Physics, Gauhati University, Guwahati-781014,India}
\date{}
\maketitle
\begin{abstract}
We use VIPT in calculating the elastic form factors and charge radii of $D,D_{s},B,B_{s}$  and $  B_{c}$  mesons in a QCD inspired potential model. For that we use linear cum Coulombic potential and opt firstly the Coulombic part as parent and then linear part as parent.The results show that charge radii and form factors are quite small for Coulombic parent case than the linear parent.Also, the analysis leads to a lower as well as upper bound on the four momentum transfer $Q^{2}$ hinting at a workable range of $Q^{2}$ within this approach which may be useful in  future for experimental analysis. Comparision of both the options prefer the linear parent.



\end{abstract}
\section{Introduction}

The potential model description in the nonperturbative regime of QCD is tremendously successful in providing both qualitative and quantitative description of the hadron spectrum and the deacy modes.On this basis,we have pursued a Non Relativisic Constituent Quark Model (NRQM)[1-4] for mesons containing a heavy (light) quark and a light(heavy) antiquark .Although non relativistic in nature ,relativistic effect is to be introduced from outside [5,6] due to the light quarks involved.Basically, the model relies on the  work of Rujula,Georgi and Glashow [7] who used a nonrelativistic treatment of potential model which was quite successful in the description of different  hadronic properties.We have solved the  Schr\"{o}dinger equation for the spin independent Fermi-Breit Hamiltonian (used in the work of Rujula et al [7]) consisting of the linear cum Coulombic potential for the ground state [2].The solution i.e. wavefunction has been obtained using different approximation methods like Dalgarno method [8] and Variationally Improved Perturbation Theory(VIPT) [9-11] which is then used in predicting the Isgur-Wise (I-W)function [2,3,4,12,13],mass,decay constant,charge radii[1] etc.We note that with the linear cum Coulomb potential of QCD we have two options in choosing parent or child(i.e.perturbation)-(i)first, we can consider the linear one as the perturbation (i.e.the Coulombic one as the parent ) and then (ii)Coulombic one as the perturbation (i.e. linear one as the parent).As we have already successfully used VIPT for both the options in the calculation of slope and curvature of I-W function for  $D,D_{s},B,B_{s}$ and $B_{c}$ mesons [12,13],so extending it for the prediction of elastic form factors and charge radii (which provide important insight on the distribution of different charge constituents of a hadron)  definitely makes a sense .

It is well known that the form factor and charge radius are  dependent on the momentum transform of the wavefunction.So,getting an appropriate wavefunction is very essential for a fruitful analysis.With the success of VIPT in the calculation of Isgur-Wise function as pointed above [12,13],one can expect a similar success here also .It is worthwhile to note that while investigating the form factor ,one must take into  account of a proper range of four momentum transfer $Q^{2}$.The  $Q^{2}$ range usually determines the applicability of perturbative QCD(pQCD) or nonperturbative(npQCD).So,an accurate selection of  $Q^{2}$ range within the nonperturbative approach is necessary which will also fall within the experimental regime.This fecilitates  a direct  comparision between theory and experiment.This has been done both theoretically and experimentally since long [14-19] for the light $\pi,K$ etc mesons.However,for the mesons which contain atleast one heavy quark,very little have been investigated theoretically [20-22].In the absence of any experimental data for them,our results may be helpful in future in the experimental set up regarding the  $Q^{2}$ range .

As far as our model is concerned, the perturbative or nonperturbative regime of QCD can be interpreted~\cite{1} through the relativistic factor `$\epsilon=1-\sqrt{1-\frac{4\alpha_{s}}{3}}$` . The reality constraint on the form factor $F\left(Q^{2}\right)$ leads to the condition $0< \epsilon<1$, where the case $\epsilon\rightarrow 0$  ( $\epsilon\rightarrow 1$) corresponds to the perturbative (nonperturbative) limit of QCD .The  $\epsilon\rightarrow 1$ limit demands large $\alpha_{s}$ or  low $Q^{2}$.So, discussing the nonperturbative effects of QCD with large  confinement parameter $b$ ,we must consider the low $Q^{2}$ limit of $\alpha_{s}$ in this model .However, we have observed in Ref[1] that  large value of $b$($=0.183 GeV^{2}$) prohibits  the use of low $Q^{2}$ compelling one to involve with small $\alpha_{s}$ which corresponds to the perturbative regime and thus  can't be accounted in this nonperturbative approach. 

We reanalyze all these observations in this approach of VIPT for both the cases -linear or confinement part as perturbation and  Coulombic part as perturbation.We will explore the possibility of incorporating significant value of $\alpha_{s}$  even with large confinement ($b=0.183GeV^{2}$). This work will also check the status of both confinement and Coulombic part as perurbation and observed the consequences regarding the usable range of $Q^{2}$ to work with in the absence of experimental data for the said mesons .The calculations are done with a fixed value of $\alpha_{s}$  from V-scheme [23-25] with large confinement effect $b=0.183 GeV^{2}$  instead of variation in both.Even with this single value of  $\alpha_{s}$ and $b$ one can draw similar conclusion regarding the effective range of $Q^{2}$. The calculated form factors are plotted graphically to  show their variation with   $Q^{2}$ for both the cases.

Basically, this work explores the possibility of improving the results for form factors and charge radii over those of Ref[1,22] with the help of VIPT.In the process, we also try to find a useful range of  $Q^{2}$ which may be workable for the experimental investigations in the later course of time.Comparision of both the options is being made to arrive at a conclusion in using VIPT.

The rest of the paper is organized as follows : section 2 contains the formalism, section 3 the result and calculation while section 4 includes the discussion and conclusion.
   
\section{Formalism}
\subsection{VIPT with Coulombic potential as parent}
\subsubsection{ Wavefunction}

 We breifly reformulate the VIPT with the expression for the  wavefunction~\cite{10,12} corrected upto the first order of $j^{th}$ state  given by :
                                                                                 \begin{equation}                                                             \psi_{j}= \psi_{j}^{(0)}+\sum_{k\neq j}\frac{\int\psi_{k}^{(0)*}H_{P^{\prime}j}^{\prime}\psi_{j}^{(0)}dv}{E_{j}^{(0)}-E_{k}^{(0)}}
   \end{equation}   

where $P^{\prime}$ is the variational parameter (which is later optimized w.r.t. energy) considered instead of physical parameter $P$.For  the Coulombic part as parent,the physical parameter is $\alpha=\frac{4\alpha_{s}}{3}$ and the optimized variational parameter is $\overline{\alpha^{\prime}}$ [12].As done in Ref[12],we consider the wave function for triple term consideration  of the summation given by equation (1) above and rewrite the equation (viz. eq(45) of Ref[12]) with the subscript $10$ ($n=1,l=0$) being replaced by $T$  :

\begin{eqnarray} 
\psi_{T}=\psi_{10}^{(0)}-A\left(1-\frac{\mu\overline{\alpha}_{10}^{\prime}r}{2}\right)e^{-\frac{\mu\overline{\alpha}_{10}^{\prime}r}{2}}+B\left(1-\frac{2\mu\overline{\alpha}_{10}^{\prime} r}{3}+\frac{2\mu^{2}\overline{\alpha}_{10}^{\prime^2}r^{2}}{27}\right)e^{-\frac{\mu\overline{\alpha}_{10}^{\prime}r}{3}}+\nonumber\\
D^{\prime}\left (\frac{1}{4}-\frac{3\mu\overline{\alpha}_{10}^{\prime} r}{16}+\frac{\mu^{2}\overline{\alpha}_{10}^{\prime^2}r^{2}}{32}-\frac{\mu^{3}\overline{\alpha}_{10}^{\prime^3}r^{3}}{8\times 96}\right)e^{-\frac{\mu\overline{\alpha}_{10}^{\prime}r}{4}}
\end{eqnarray}
where the different parameters are given by:

\begin{equation}
c_{1}^{\prime}=\frac{\mu\overline{\alpha}_{10}^{\prime}}{\pi^{\frac{1}{3}}}
\end{equation}

\begin{equation}
A=\frac{4\sqrt{\mu}}{3\sqrt{\pi}\left(\overline{\alpha}_{10}^{\prime}\right)^{\frac{1}{2}}}[\frac{4\mu\overline{\alpha}_{10}^{\prime}\left(\alpha-\overline{\alpha}_{10}^{\prime}\right)}{27}-\frac{32b}{81\mu\overline{\alpha}_{10}^{\prime}}]
\end{equation}
\begin{equation}
B=\frac{\sqrt{\mu}}{\sqrt{\pi}\left(\overline{\alpha}_{10}^{\prime}\right)^{\frac{1}{2}}}[\frac{3\mu\overline{\alpha}_{10}^{\prime}\left(\alpha-\overline{\alpha}_{10}^{\prime}\right)}{64}-\frac{27b}{256\mu\overline{\alpha}_{10}^{\prime}}]
\end{equation}
and
\begin{equation}
D^{\prime} =\frac{\left(\mu\overline{\alpha}_{10}^{\prime}\right)^{\frac{3}{2}}}{\sqrt{\pi}}[\frac{36\left(\alpha-\overline{\alpha}_{10}^{\prime}\right)}{15625\overline{\alpha}_{10}^{\prime}}-\frac{384b}{78125\mu^{2}\overline{\alpha}_{10}^{\prime^3}}]
\end{equation} 
 The wavefunctions for single and double term consideration can be obtained by putting $B=D^{\prime}=0$ and $D^{\prime}=0$ respectively in the same equation (2).

However we will consider the relativistic version ($\epsilon\ne0$) of the above wave function  viz.~\cite{12}

\begin{equation}
\psi_{T,Rel}\left(r\right)=\psi_{T}\left(r\mu\overline{\alpha}_{10}^{\prime}\right)^{-\epsilon}
\end{equation}
The relativistic factor $\epsilon$ is given by ~\cite{1}:
\begin{equation}
\epsilon =1-\sqrt{1-\frac{4\alpha_{s}}{3}}
\end{equation}
\subsubsection{The elastic charge form factor and charge radii}
The form factor can be expressed as ~\cite{26}:
\begin{equation}                                                          
   eF\left(Q^{2}\right)= \sum\frac{e_{i}}{Q_{i}}\int_{0}^{+\infty}  r\left|\psi_{T,Rel}\left(r\right)\right|^{2}\sin Q_{i}r dr
     \end{equation}  
where
\begin{equation}
Q_{i}=\frac{\sum_{j\neq i}m_{j}Q}{\sum m_{i}}
\end{equation}
Putting (2) and (7) in (9) we get the form factor as :
\begin{equation}
 eF\left(Q^{2}\right)=\sum e_{i}N_{3}^{\prime^{2}}\Gamma(3-2\epsilon)\left[q_{1}+q_{2}+q_{3}+q_{4}+q_{5}+q_{6}+q_{7}+q_{8}+q_{9}+q_{10}\right]
\end{equation}
where $N_{3}^{\prime^{2}}$ is the same normalization constant as appeared in equation(54)of Ref[12] and   the different $q_{i}(Q_{i})$ s($i=1,2,...,10$) are defined in the Appendix.

The charge radius is derived as ~\cite{1} :
\begin{eqnarray}
<r^{2}>&=&-\left. \frac{d\left( eF\left(Q^{2}\right)\right)}{d Q^{2}}\right.|_{Q^{2}=0}\\\nonumber &=& N_{3}^{\prime^{2}}\Gamma(3-2\epsilon)[r_{1}+r_{2}+r_{3}+r_{4}+r_{5}+r_{6}+r_{7}+r_{8}\\&+&r_{9}+r_{10}]
\end{eqnarray}
where the different $r_{i}$ s ($i=1,2,...,10$) are defined in the Appendix.

\subsubsection{Status of linear potential as perturbation}
The momentum transform of equation (7) is ~\cite{27,28}:

\begin{eqnarray}
\psi_{T,Rel}\left(Q^{2}\right)&=& \sum\frac{e_{i}}{Q_{i}}\sqrt{\frac{2}{\pi}}\int_{0}^{+\infty}  r\psi_{T,Rel}\left(r\right)sin Q_{i}r dr\\&=&\sum e_{i}N_{3}^{\prime}\sqrt{\frac{2}{\pi}}\Gamma(3-2\epsilon)\left[p_{1}-p_{2}+p_{3}+p_{4}\right]
\end{eqnarray}
The $p_{i}$ s which depend on $Q_{i}^{2},\epsilon$ etc are given in the Appendix.

If linear potential is treated as perturbation then from equation(15) above the following inequality must be preserved:

\begin{equation}
p_{1}>p_{2}-p_{3}-p_{4}
\end{equation}
This inequality leads to a lower limit of $Q^{2}$ namely $Q_{0}^{2}$ ~\cite{1} above which one have to use the values of  $Q^{2}$.The $Q_{0}^{2}$ is determined from the condition:

\begin{equation}
p_{1}=p_{2}-p_{3}-p_{4}
\end{equation}
   Due to the  quark mass dependence ,$Q_{0}^{2}$ s have different values and they are shown in table 3.In the Dalgarno method approach ~\cite{1}, the lower limits $Q_{0}^{2}$ were large and the formalism failed to account for large confinement effect ($b=0.183GeV^{2}$)in the nonperturbative QCD regime where $\alpha_{s}$ values were taken to be large.Only in the limit $b\rightarrow{0}$ ,the $Q_{0}^{2}$ values were lowered and the  formalism worked for low  $Q^{2}$ range ~\cite{1}.In this method of VIPT,the values of  $Q_{0}^{2}$ are shown to be quite  small even with large confinement effect ($b=0.183GeV^{2}$) enabling us to work in the nonperturbative QCD regime with large  $\alpha_{s}$.

We also note that for single term consideration only $p_{2}$ exists on the RHS of inequality and for double term both $p_{2}$ and $p_{3}$  exist.We have also recorded the values of $Q_{0}^{2}$ for single and double term consideration in table 3.  
\subsection{VIPT with linear  potential as parent}
\subsubsection{ Wavefunction}
As pointed in Ref[10,13] linear parent gives rise to Airy functions.The physical parameter is $b$ and the optimized variational parameter is $\overline{b^{\prime}}$ .We reproduce the analogous wavefunction  in this case also for three term consideration of eq.(1) as was for the Coulombic parent: 

\begin{eqnarray}
\psi_{T}=N^{\prime\prime}[\psi^{(0)}+\nonumber\frac{\left(2\mu\right)^{\frac{1}{3}}}{\left(\rho_{02}-\rho_{01}\right)\overline{b}^{\prime^{\frac{2}{3}}}}\left(\left(b-\overline{b}^{\prime}\right)<r>_{2,1}-\alpha<\frac{1}{r}>_{2,1}\right)\psi_{20}\left(r\right)+
\\\nonumber \frac{\left(2\mu\right)^{\frac{1}{3}}}{\left(\rho_{03}-\rho_{01}\right)\overline{b}^{\prime^{\frac{2}{3}}}}\left(\left(b-\overline{b}^{\prime}\right)<r>_{3,1}-\alpha<\frac{1}{r}>_{3,1}\right)\psi_{30}\left(r\right)\\
+\frac{\left(2\mu\right)^{\frac{1}{3}}}{\left(\rho_{04}-\rho_{01}\right)\overline{b}^{\prime^{\frac{2}{3}}}}\left(\left(b-\overline{b}^{\prime}\right)<r>_{4,1}-\alpha<\frac{1}{r}>_{4,1}\right)\psi_{40}\left(r\right)]
\end{eqnarray}
where $ N^{\prime\prime}$ is the normalization constant as appeared in equation(25) of Ref[13].We note that for single (double) term consideration of equation(1)  the third and fourth term ( fourth term) is dropped from the equation (18) and normalization constants also changes to different one (Eq.17 and 21 of Ref [13]).

For this case also we take the relativistic version of the above wavefunction :
\begin{equation}
\psi_{T,Rel}=\psi_{T}\left(r\mu\overline{\alpha}_{10}^{\prime}\right)^{-\epsilon}
\end{equation}
The zeros of the Airy function $\rho_{0n}$ is given by eq.(11) of Ref[13] as:
\begin{equation}
\rho_{on}=-\left[\frac{3\pi(4n-1)}{8}\right]^{\frac{2}{3}}
\end{equation}
and 
\begin{equation}
<r^{k}>_{n,n^{\prime}}=N_{n}N_{n^{\prime}}\int_{0}^{+\infty}r^{k}Ai\left((2\mu \overline{b}^{\prime})^{\frac{1}{3}}r -\rho_{0n}\right)Ai\left((2\mu \overline{b}^{\prime})^{\frac{1}{3}}r -\rho_{0n^{\prime}}\right)dr
\end{equation}
where $N_{n},N_{n^{\prime}}$ are the normalization constants for $n$ and $n^{\prime}$ states respectively etc.

Like the expressions we have adopted the same values of $b,\overline{b}^{\prime},\rho_{0n}$ from Ref[13](see tables 1,2 there). 
\subsubsection{The elastic charge form factor and charge radii}
Putting (18) and (19) in the definition of form factor (9) given above, the form factor is found to be :
\begin{equation}
 eF\left(Q^{2}\right)=\sum e_{i} N^{\prime\prime^{2}}\left[C-C^{\prime}\frac{Q_{i}^{2}}{6}\right]
\end{equation}
The coefficients $C,C^{\prime}$s are given in table 4.They are of course different for single ,double or more than two term consideration.
Numerical integrations are done in getting the above result.

The corresponding charge radius is obtained by using equation (18) and (19) in(12) which are recorded in table 4.

\subsubsection{Status of Coulombic potential as perturbation}
The momentum transform of (19) is:

  \begin{eqnarray}
\psi_{T,Rel}\left(Q^{2}\right)&=& \sum\frac{e_{i}}{Q_{i}}\sqrt{\frac{2}{\pi}}\int_{0}^{+\infty}  r\psi_{T,Rel}\left(r\right)sin Q_{i}r dr\\\nonumber&=&\sum e_{i}N_{3}^{\prime}\sqrt{\frac{2}{\pi}}\Gamma(3-2\epsilon)[p_{1}^{\prime}+p_{2}^{\prime}+p_{3}^{\prime}+p_{4}^{\prime}\\&-&\frac{Q_{i}^{2}}{6}\left(p_{5}^{\prime}+p_{6}^{\prime}+p_{7}^{\prime}+p_{8}^{\prime}\right)]
\end{eqnarray}
The $p_{i}^{\prime}$ s ($i=1,2,..,8$) are given in the Appendix.

If Coulombic  potential is treated as perturbation then from equation(24) above the following inequality must be preserved:

\begin{equation}
p_{1}^{\prime}+p_{2}^{\prime}+p_{3}^{\prime}+p_{4}^{\prime}>\frac{Q_{i}^{2}}{6}\left(p_{5}^{\prime}+p_{6}^{\prime}+p_{7}^{\prime}+p_{8}^{\prime}\right)
\end{equation}

This inequality leads to a upper limit of $Q^{2}$ namely $Q_{0}^{2}$  below  which one have to use the values of  $Q^{2}$.The $Q_{0}^{2}$ is determined from the condition:

\begin{equation}
p_{1}^{\prime}+p_{2}^{\prime}+p_{3}^{\prime}+p_{4}^{\prime}=\frac{Q_{i}^{2}}{6}\left(p_{5}^{\prime}+p_{6}^{\prime}+p_{7}^{\prime}+p_{8}^{\prime}\right)
\end{equation}
The different values of upper limit $Q_{0}^{2}$ s are shown in table 5.The corresponding values for single and double term consideration are also shown.

 \begin{figure}[h]
\subfigure{
\includegraphics[width=0.33\textwidth,angle=270]{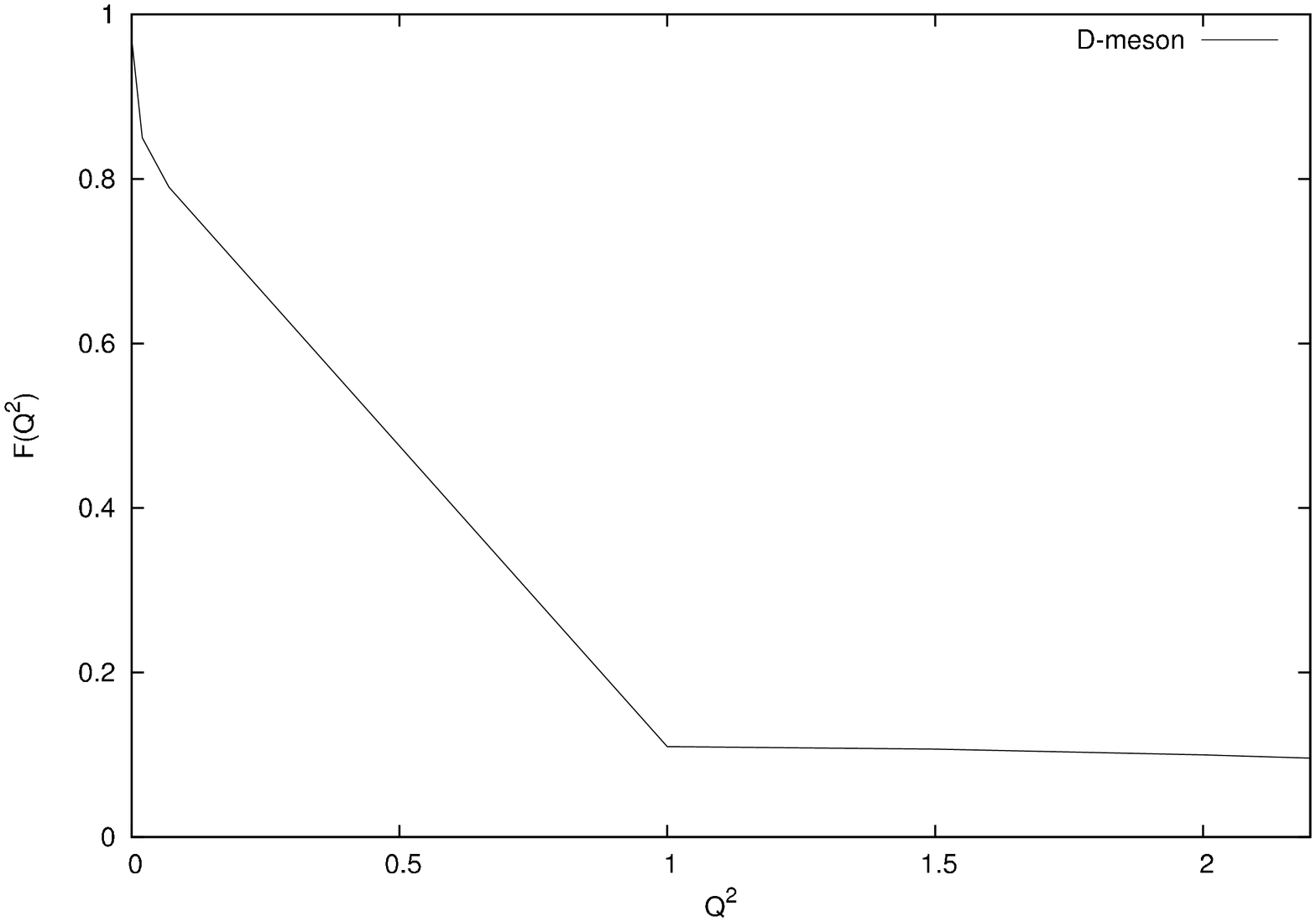}
          }
\subfigure{
\includegraphics[width=0.33\textwidth,angle=270]{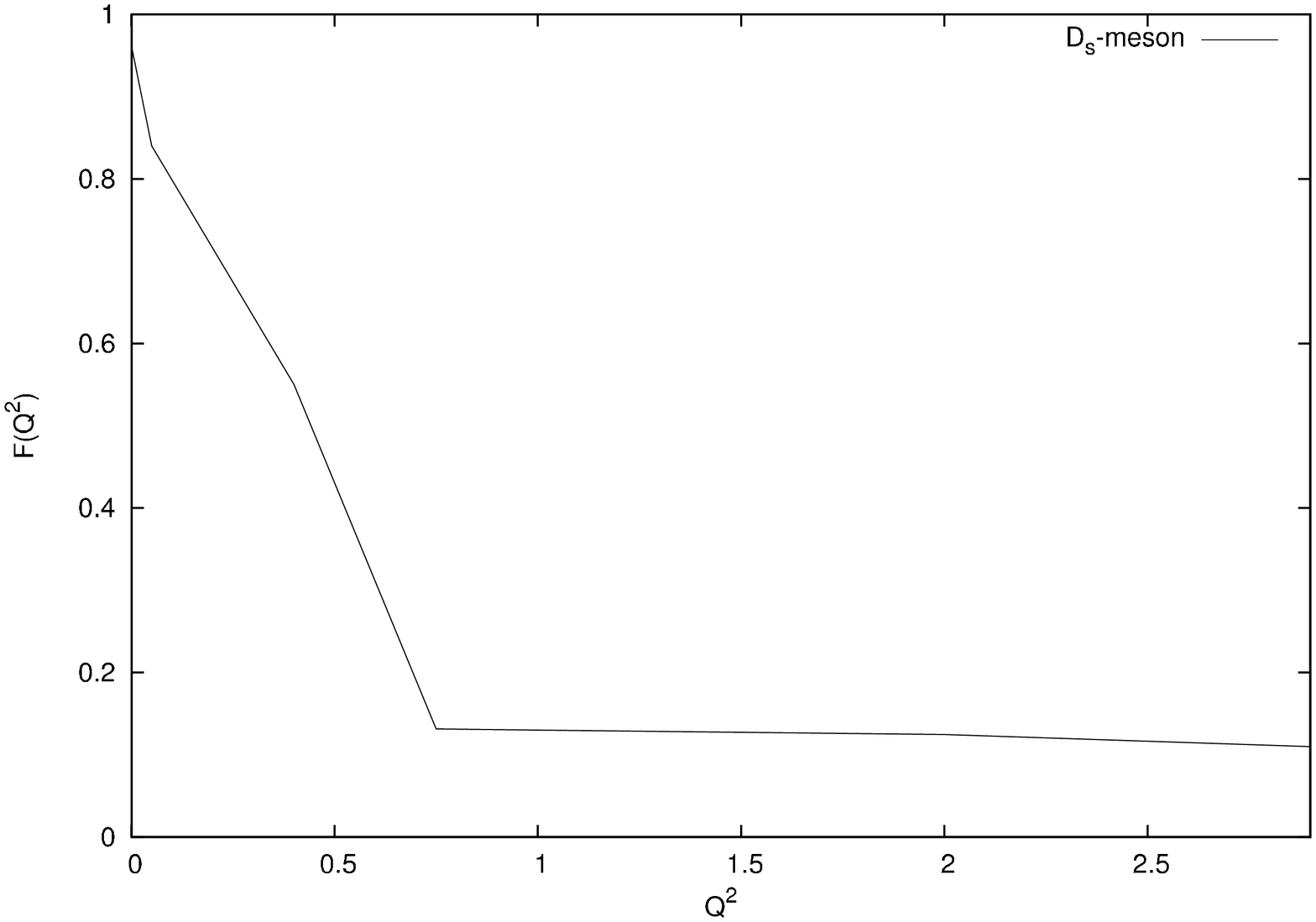}
}
\subfigure{
\includegraphics[width=0.33\textwidth,angle=270]{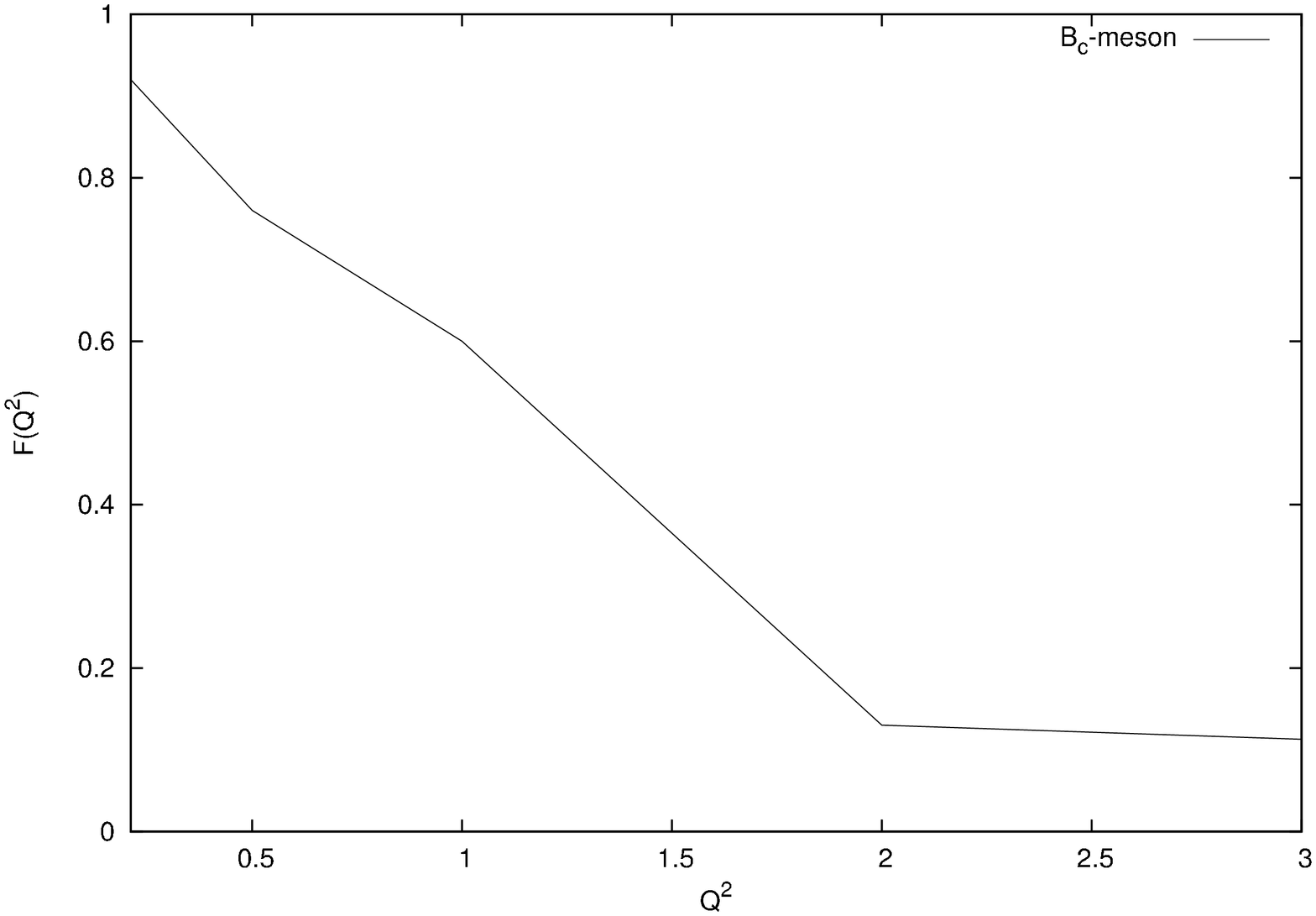}
}
\caption{Variation of $eF(Q^{2})$ vs $Q^{2}$ for  $ D$ ,$D_{s}$ and  $B_{c}$-meson with Coulombic parent .}
\end{figure}
\begin{figure}[h]
\subfigure{
\includegraphics[width=0.33\textwidth,angle=270]{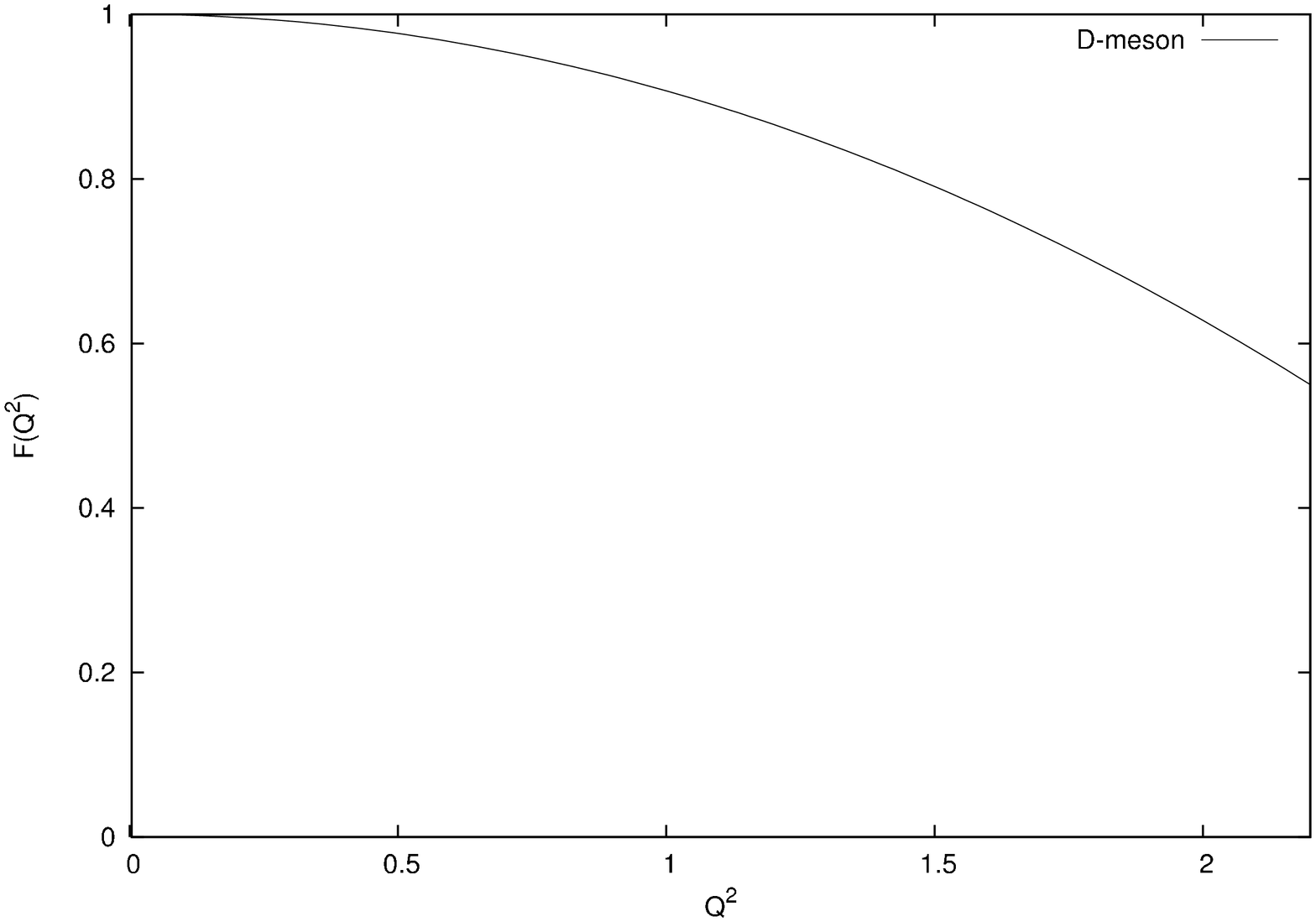}}
\subfigure{
\includegraphics[width=0.33\textwidth,angle=270]{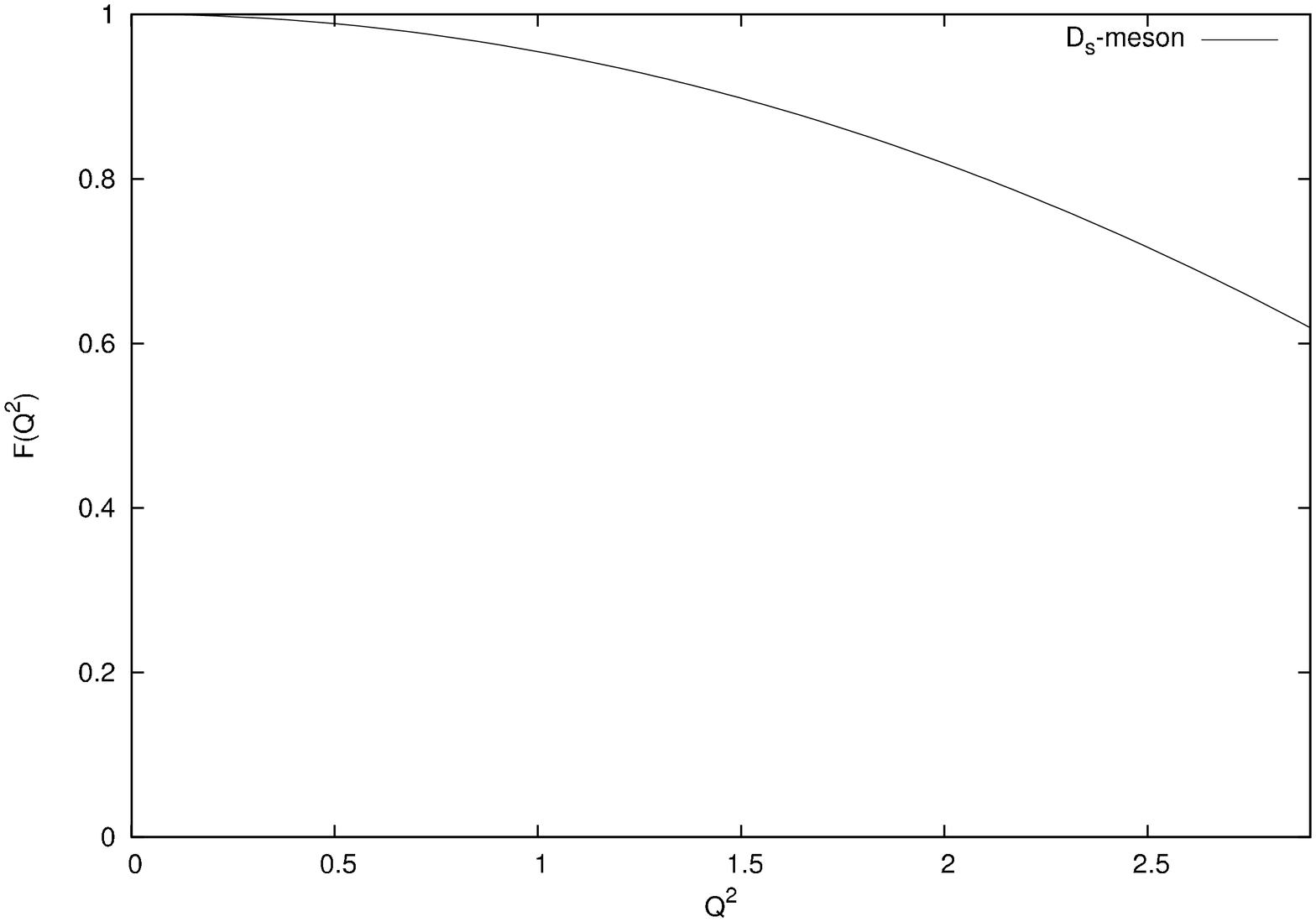}}
\subfigure{
\includegraphics[width=0.33\textwidth,angle=270]{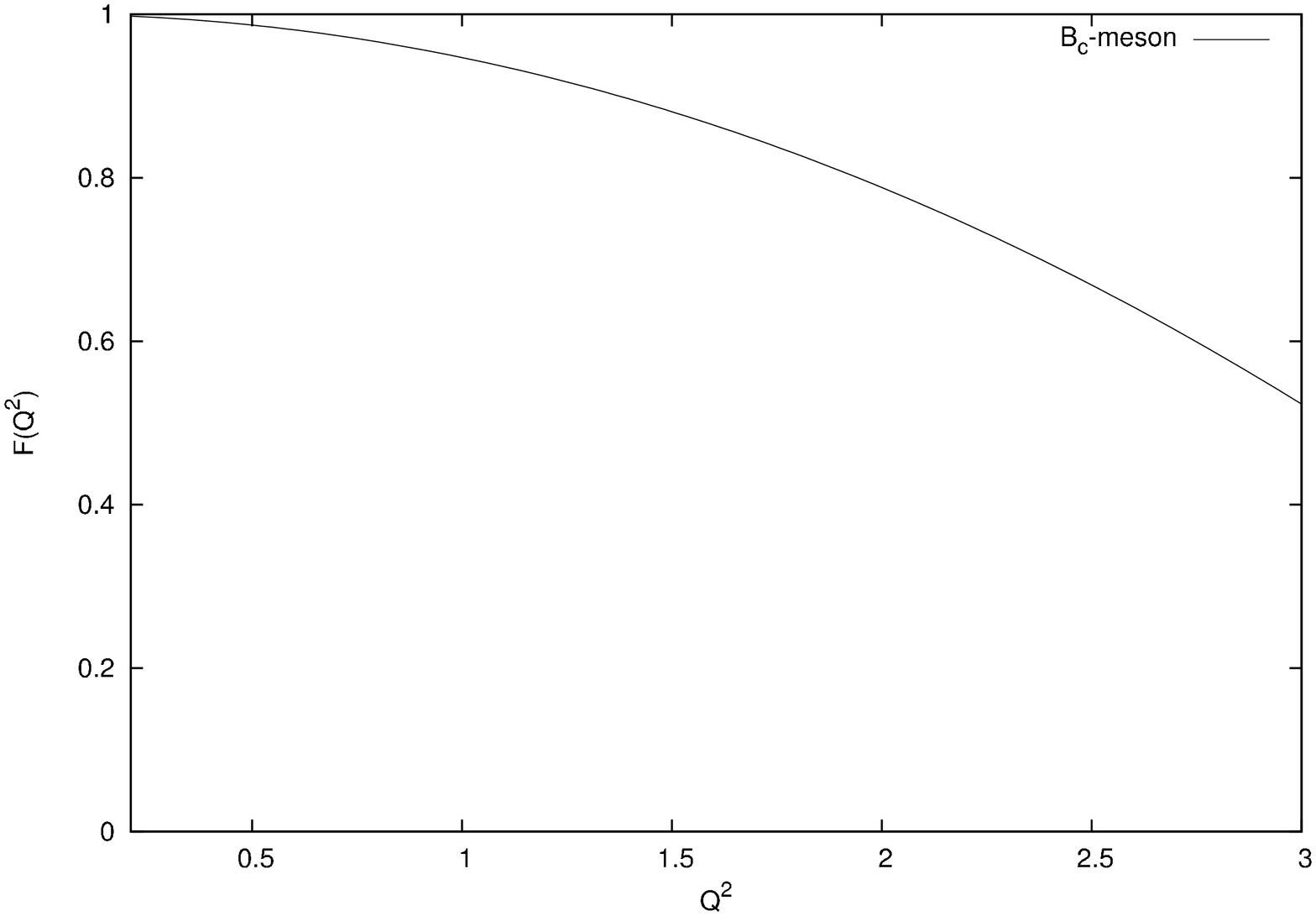}}
\caption{Variation of $eF(Q^{2})$ vs $Q^{2}$ for $ D$ ,$D_{s}$ and  $B_{c}$-meson with linear parent .}
\end{figure}

 \section{Calculation and Results}
We have listed $\overline{b}^{\prime}$ in table 1 while the lower and upper limit of $Q_{0}^{2}$ for single,double and triple term consideration are given in tables 3 and 5.
In table 2 , we record the charge radii for single,double and triple term consideration for  Coulombic potential as parent; whereas the same is recorded for linear potential as parent in table 4.The infinite mass consideration shown by the subscript $\infty$  is also included for triple (single) term consideration for Coulombic  (linear) parent.The table 6 shows charge radii of different mesons obtained from  other models and data.
The  $\alpha_{s}$ values are taken from the  $V$-scheme ~\cite{23,24,25} and the integrations are done numerically for all these calculations.
The graphs show the variation $eF(Q^{2})$ vs  $Q^{2}$ for $D,D_{s}$ and $B_{c}$  mesons for both the options.
\newpage
\begin{table}[ht]
\begin{center}
\caption{Values of $\overline{b}^{\prime}$ with $b=0.183GeV^{2}$ for relativistic case only.The values are  the same as recorded in table 2 of Ref[13]}
\vspace{0.2in}
\begin{tabular}{c c c c}\hline
Mesons&Reduced mass $\mu$&$\alpha=\frac{4\alpha_{s}}{3}$&$\overline{b}^{\prime}$with
 relativistic effect\\\hline
$D$&0.2761&0.924&16.24\\
$D_{s}$&0.368248&0.924&19.8\\
$B$&0.31464&0.348&5.587\\
$B_{s}$&0.4401&0.348&5.954\\
$B_{c}$&1.1803&0.348&8.103\\\hline
\end{tabular}
\end{center}
\end{table}

\begin{table}[ht]
\begin{center}
\caption{Values of charge radii for different mesons with Coulombic parent for single,double and triple terms in eq.(1).The subscripts `$S,D,T$' correspond to single,double and triple terms respectively whereas `$ F$' means finite mass consideration.The infinite mass limit ( subscript $\infty$ is used) is  shown for the triple term alone . }
\vspace{0.2in}
\begin{tabular}{c c c c c c c c}\hline
Meson &$D^{0}$&$D^{+}$&$D_{s}^{+}$&$B^{+}$&$B^{0}$&$B_{s}^{0}$&$B_{c}^{+}$\\
 & & & & & & &\\\hline
$<r_{S,F}^{2}>^{\frac{1}{2}}in fm$&-0.121&0.115&0.11&0.2545&-0.1822&-0.168&0.108\\
$<r_{D,F}^{2}>^{\frac{1}{2}}in fm$&-0.119&0.112&0.106&0.2512&-0.1788&-0.164&0.105\\
$<r_{T,F}^{2}>^{\frac{1}{2}}in fm$&-0.118&0.109&0.101&0.2464&-0.1736&-0.158&0.1034\\
$<r_{T,\infty}^{2}>^{\frac{1}{2}}in fm$&-0.131&0.12&0.113&0.263&-0.186&-0.1742&0.1325\\\hline

\end{tabular}
\end{center}
\end{table}

\begin{table}[ht]

\begin{center}
\caption{Values of lower limit of four momentum transfer  $Q_{0}^{2}$ with Coulombic parent taking single,double and triple terms in eq.(1).We have to use  $Q^{2}$ values above these. }
\vspace{0.2in}
\begin{tabular}{c c c c c c c c}\hline
Meson&$D^{+}$&$D^{-}$&$D_{s}^{+}$&$B^{+}$&$B^{0}$&$B_{s}^{0}$&$B_{c}^{+}$\\\hline
$Q_{0,S}^{2}$&0.0004&0.0004&0.001&0.053&0.053&0.075&0.211\\
$Q_{0,D}^{2}$&0.00036&0.00036&0.0009&0.052&0.052&0.072&0.209\\
$Q_{0,T}^{2}$&0.0003&0.0003&0.0007&0.05&0.05&0.07&0.205\\\hline

\end{tabular}
\end{center}
\end{table}
\newpage

\begin{table}[h]
\begin{center}
\caption{Values of charge radii for different mesons with linear parent for single,double and triple terms in  eq.(1) . The subscripts  `$S,D,T$' correspond to single,double and triple terms  respectively whereas `$F$' means finite mass consideration .The infinite mass limit ( subscript $\infty$ is used) is shown for the single term alone. }
\vspace{0.25in}
\begin{tabular}{c c c c c c c c}\hline
Meson&$D^{0}$&$D^{+}$&$D_{s}^{+}$&$B^{+}$&$B^{0}$&$B_{s}^{0}$&$B_{c}^{+}$\\\hline
$C_{S}$&8.24&8.24&5.916&14.22&14.22&10.91&4.50\\
$C^{\prime}_{S}$&2.266&2.266&1.173&7.71&7.71&4.52&0.78\\
$<r_{S,F}^{2}>^{\frac{1}{2}}in fm$&-0.197&0.1494&0.104&0.425&-0.2996&-0.2227&0.1125\\
$C_{D}$&13.1&13.1&8.1&26.7&26.7&16.5&5.2\\
$C^{\prime}_{D}$&2.69&2.69&1.67&9.89&9.89&5.7&1.1\\
$<r_{D,F}^{2}>^{\frac{1}{2}}in fm$&-0.21&0.161&0.121&0.473&-0.336&-0.2489&0.127\\
$C_{T}$&18.14&18.14&11.18&43.09&43.09&25.56&6.636\\
$C^{\prime}_{T}$&3.365&3.365&2.199&13.14&13.14&7.655&1.363\\
$<r_{T,F}^{2}>^{\frac{1}{2}}in fm$&-0.24&0.182&0.143&0.555&-0.391&-0.289&0.148\\
$<r_{S,\infty}^{2}>^{\frac{1}{2}}in fm$&-0.246&0.174&0.125&0.453&-0.32&-0.246&0.144\\\hline
\end{tabular}
\end{center}
\end{table}

\vspace{1.5in}
\begin{table}[ht]
\begin{center}
\caption{Values of upper limit of four momentum transfer  $Q_{0}^{2}$ with linear parent taking single,double and triple terms in eq.(1).We have to use  $Q^{2}$ values lower than  these. }
\vspace{0.2in}
\begin{tabular}{c c c c c c c c}\hline
Meson&$D^{+}$&$D^{-}$&$D_{s}^{+}$&$B^{+}$&$B^{0}$&$B_{s}^{0}$&$B_{c}^{+}$\\\hline
$Q_{0,S}^{2}$&2.297&2.297&2.92&1.43&1.43&1.676&3.11\\
$Q_{0,D}^{2}$&2.1&2.1&2.67&1.31&1.31&1.56&2.89\\
$Q_{0,T}^{2}$&1.88&1.88&2.387&1.177&1.177&1.386&2.55\\\hline
\end{tabular}
\end{center}
\end{table}

\newpage
\begin{table}[ht]
\begin{center}
\caption{Prediction of $<r^{2}>^{\frac{1}{2}}in fm$ for finite and infinite mass consideration in other models.The subscript `$F$' (`$\infty$' ) means finite (infinite) mass limit.}

\begin{tabular}{c c c c c c c c}\hline
Meson&$D^{0}$&$D^{+}$&$D_{s}^{+}$&$B^{+}$/$B^{-}$&$B^{0}$&$B_{s}^{0}$/$\overline{B_{s}^{0}}$&$B_{c}^{+}$/$B_{c}^{-}$\\\hline
$<r^{2}_{F}>^{\frac{1}{2}}$[9]&...&0.506&0.491&0.258($B^{-}$)&...&0.256($\overline{B_{s}^{0}}$)&0.236($B_{c}^{-}$)\\
$<r^{2}_{F}>^{\frac{1}{2}}$[10]&-0.551&0.43&0.352&0.612($B^{+}$)&-0.432&-0.345($B_{s}^{0}$)&0.207($B_{c}^{+}$)\\
$<r^{2}_{\infty}>^{\frac{1}{2}}$[10]&-0.704&0.498&0.425&0.704($B^{+}$)&-0.498&-0.425($B_{s}^{0}$)&...($B_{c}^{+}$)\\
$<r^{2}_{F}>^{\frac{1}{2}}$[11]&-0.484&0.366&0.355&1.72($B^{+}$)&-1.21&-1.17($B_{s}^{0}$)&1.43($B_{c}^{+}$)\\
$<r^{2}_{\infty}>^{\frac{1}{2}}$[11]&-0.6025&0.427&0.427&1.836($B^{+}$)&-1.29&-1.29($B_{s}^{0}$)&1.84($B_{c}^{+}$)\\\hline
\end{tabular}
\end{center}
\end{table}

\section{Discussion and Conclusion}
We have analyzed elastic form factors and charge radii in a QCD inspired potential model using VIPT under two scenarios-Coulombic potential and linear potential as parent . 

We summarize our achievements below:

I. The form factor $eF(Q^{2})$ decreases  with the increase of $Q^{2}$ (as it should) for both the scenarios.

II. The form factor is either very small (for $D$-sector mesons)or small ( for $B$-sector mesons)with Coulombic parent as compared to those with linear parent.

The charge radii is also observed to be smaller with Coulombic parent as compared to linear parent.

III. We use a fixed set of values for $\alpha_{s}$ under $V$-scheme[23-25] in the calculation , for example it is $0.693$ for the  $D,D_{s}$ mesons which is larger than the value $0.261$ for the $B,B_{s},B_{c}$ mesons.This consideration directly results in the unexpected  smaller values of  charge radii for $D,D_{s}$ mesons as compared to the $B,B_{s},B_{c}$ mesons.Larger $\alpha_{s}$ values are responsible for smaller charge radius.

IV. While checking the status of confinement as perturbation i.e. Coulombic parent (or Coulombic part as perturbation i.e. linear parent) we end up with a lower (or upper) limit on $Q^{2}$ .This allows us a useful range of $Q^{2}$  to show the variation of form factor which is shown in fig 1 and 2.

V. In the present analysis, even with large $b$,the lower limit of $Q^{2}$(for linear perturbation)  are really small as shown in table 3 for fixed $\alpha_{s}$.We have seen that for $\alpha_{s}=0.693$ , the lower limit of $Q^{2}$ for $D,D_{s}$ mesons are respectively $0.0003,0.0007$, whereas with  $\alpha_{s}=0.261$ ,the lower limit of   $Q^{2}$ for $B,B_{s},B_{c}$ mesons are respectively $0.05,0.07,0.205 $ .These values for $B,B_{s},B_{c}$ mesons will be lowered if we put  $\alpha_{s}>0.261$. This is clearly advantageous over the Dalgarno method with linear perturbation as done in Ref[1] where the formalism broke down for large $b$.Thus, this approach allows a large value of  $\alpha_{s}(Q^{2})$ in the limit $Q^{2}\rightarrow{0}$ even with large confinement , an important feature absent in Ref[1].

VI.  The Coulombic perturbation leads to an upper limit of  $Q^{2}$ in this case (table 5) which allows us to use any value of  $\alpha_{s}(Q^{2})$ in the limit $Q^{2}\rightarrow{0}$.

VII. Further, if we look at Eq.(1), consideration of different terms leads to different  charge radii and the limiting values of  $Q^{2}$  for both the cases. The charge radii and the lower limit of  $Q^{2}$ decrease with more terms for the linear part as perturbation  (i.e. Coulombic parent) whereas the charge radii increase  and  upper limit of  $Q^{2}$ decreases for the Coulombic part as perturbation (i.e.  linear parent).

VIII. The infinite mass consideration in this work shows that the charge radii are larger than those for finite mass consideration  to agree well with other models (table 6).  

The above list  as a whole suggests success of VIPT over the Dalgarno method ~\cite{1,22} as far as large confinement and limiting values of  $Q^{2}$ are concerned. The difference in the values of form factors and charge radii for both the cases  may be attributed to the use of same  $\alpha_{s}$ (i.e. $Q^{2}$) under $V$-scheme for both the scenarios as the Coulombic potential is dominant for large  $Q^{2}$ (i.e.low $r$)and the linear potential in the low  $Q^{2}$ (i.e. large $r$)regime.We must note that we have used the low $Q^{2}$ assumption in the calculation of form factors and this clearly effects the upper limit of $Q^{2}$  corresponding to the validity of Coulombic perturbation (i.e. linear parent).The larger value of $\alpha_{s}$ for $D$-sector as compared to $B$-sector is also another point to take account of this .Although, the linear parent has shown more flexibility and hence is the better option than Coulombic parent in VIPT,but it has used terms up to a particular order in`$r$' in the integration involved with Airy function (which is an infinite series).This may lead to loss of certain information as far as physics is concerned. In the absence of any experimental results for these mesons , it is quite difficult to make a direct conclusion  but there is clear indication that one must be careful in choosing the parameter $\alpha_{s}(Q^{2})$ as well as the confinement parameter in the calculation of form factor and charge radius within the QCD framework.

The above discussion led to the conclusion that there is scope to use this approach in the study of meson decays.The lower and upper limit on  $Q^{2}$ (i.e. range of  $Q^{2}$) in this analysis may be useful in the experimental set up to investigate cross-section and form factor in future  for these mesons.Further, from the model specific values of form factors and charge radii  the method  allows  to investigate the behaviour of $\alpha_{s}$ w.r.t  $Q^{2}$  in the nonperturbative regime of QCD.

\vspace{1in}
\appendix

\textbf{Expressions for $q_{i}$ s} :

 \begin{equation}
q_{1}=\frac{c_{1}^{\prime^{2}}}{(4\mu^{2}\overline{\alpha}^{\prime^{2}}+Q_{i}^{2})^{(1-\epsilon)}}
\end{equation}

 \begin{equation}
q_{2}=A^{2}\left[\frac{1}{(\mu^{2}\overline{\alpha}^{\prime^{2}}+Q_{i}^{2})^{(1-\epsilon)}}-\frac{(3-2\epsilon)\mu\overline{\alpha}^{\prime}}{(\mu^{2}\overline{\alpha}^{\prime^{2}}+Q_{i}^{2})^{(1.5-\epsilon)}}+\frac{(4-2\epsilon)(3-2\epsilon)\mu^{2}\overline{\alpha}^{\prime^{2}}}{(\mu^{2}\overline{\alpha}^{\prime^{2}}+Q_{i}^{2})^{(2-\epsilon)}}\right]
\end{equation}

 \begin{eqnarray}
q_{3}=B^{2}[\nonumber\frac{1}{(\frac{\mu^{2}\overline{\alpha}^{\prime^{2}}}{9}+Q_{i}^{2})^{(1-\epsilon)}}-\frac{4(3-2\epsilon)\mu\overline{\alpha}^{\prime}}{3(\frac{\mu^{2}\overline{\alpha}^{\prime^{2}}}{9}+Q_{i}^{2})^{(1.5-\epsilon)}}&+&\\
\nonumber\frac{16(4-2\epsilon)(3-2\epsilon)\mu^{2}\overline{\alpha}^{\prime^{2}}}{27(\frac{\mu^{2}\overline{\alpha}^{\prime^{2}}}{9}+Q_{i}^{2})^{(2-\epsilon)}}-\frac{8(5-2\epsilon)(4-2\epsilon)(3-2\epsilon)\mu^{3}\overline{\alpha}^{\prime^{3}}}{81(\frac{\mu^{2}\overline{\alpha}^{\prime^{2}}}{9}+Q_{i}^{2})^{(2.5-\epsilon)}}&+&\\\frac{4(6-2\epsilon)(5-2\epsilon)(4-2\epsilon)(3-2\epsilon)\mu^{4}\overline{\alpha}^{\prime^{4}}}{729(\frac{\mu^{2}\overline{\alpha}^{\prime^{2}}}{9}+Q_{i}^{2})^{(3-\epsilon)}}]
\end{eqnarray}
\begin{eqnarray}
q_{4}=D^{\prime^{2}}[\nonumber\frac{1}{16(\frac{\mu^{2}\overline{\alpha}^{\prime^{2}}}{4}+Q_{i}^{2})^{(1-\epsilon)}}-\frac{3(3-2\epsilon)\mu\overline{\alpha}^{\prime}}{32(\frac{\mu^{2}\overline{\alpha}^{\prime^{2}}}{4}+Q_{i}^{2})^{(1.5-\epsilon)}}&+&\\
\nonumber\frac{17(4-2\epsilon)(3-2\epsilon)\mu^{2}\overline{\alpha}^{\prime^{2}}}{256(\frac{\mu^{2}\overline{\alpha}^{\prime^{2}}}{4}+Q_{i}^{2})^{(2-\epsilon)}}-\frac{19(5-2\epsilon)(4-2\epsilon)(3-2\epsilon)\mu^{3}\overline{\alpha}^{\prime^{3}}}{1536(\frac{\mu^{2}\overline{\alpha}^{\prime^{2}}}{4}+Q_{i}^{2})^{(2.5-\epsilon)}}&+&\\\nonumber\frac{7(6-2\epsilon)(5-2\epsilon)(4-2\epsilon)(3-2\epsilon)\mu^{4}\overline{\alpha}^{\prime^{4}}}{6144(\frac{\mu^{2}\overline{\alpha}^{\prime^{2}}}{9}+Q_{i}^{2})^{(3-\epsilon)}}&-&\\\frac{(7-2\epsilon)(6-2\epsilon)(5-2\epsilon)(4-2\epsilon)(3-2\epsilon)\mu^{5}\overline{\alpha}^{\prime^{5}}}{12288(\frac{\mu^{2}\overline{\alpha}^{\prime^{2}}}{4}+Q_{i}^{2})^{(3.5-\epsilon)}}]
\end{eqnarray}
\begin{equation}
q_{5}=2c_{1}^{\prime}A\left[\frac{1}{(\frac{9\mu^{2}\overline{\alpha}^{\prime^{2}}}{4}+Q_{i}^{2})^{(1-\epsilon)}}-\frac{(3-2\epsilon)\mu\overline{\alpha}^{\prime}}{2(\frac{9\mu^{2}\overline{\alpha}^{\prime^{2}}}{4}+Q_{i}^{2})^{(1.5-\epsilon)}}\right]
\end{equation}

 \begin{equation}
q_{6}=2c_{1}^{\prime}B\left[\frac{1}{(\frac{16\mu^{2}\overline{\alpha}^{\prime^{2}}}{9}+Q_{i}^{2})^{(1-\epsilon)}}-\frac{2(3-2\epsilon)\mu\overline{\alpha}^{\prime}}{3(\frac{16\mu^{2}\overline{\alpha}^{\prime^{2}}}{9}+Q_{i}^{2})^{(1.5-\epsilon)}}+\frac{2(4-2\epsilon)(3-2\epsilon)\mu^{2}\overline{\alpha}^{\prime^{2}}}{27(\frac{16\mu^{2}\overline{\alpha}^{\prime^{2}}}{9}+Q_{i}^{2})^{(2-\epsilon)}}\right]
\end{equation}

 \begin{eqnarray}
q_{7}=2c_{1}^{\prime}D^{\prime}[\nonumber\frac{1}{4(\frac{25\mu^{2}\overline{\alpha}^{\prime^{2}}}{16}+Q_{i}^{2})^{(1-\epsilon)}}-\frac{3(3-2\epsilon)\mu\overline{\alpha}^{\prime}}{16(\frac{25\mu^{2}\overline{\alpha}^{\prime^{2}}}{16}+Q_{i}^{2})^{(1.5-\epsilon)}}&+&\\
\frac{(4-2\epsilon)(3-2\epsilon)\mu^{2}\overline{\alpha}^{\prime^{2}}}{32(\frac{25\mu^{2}\overline{\alpha}^{\prime^{2}}}{16}+Q_{i}^{2})^{(2-\epsilon)}}-\frac{(5-2\epsilon)(4-2\epsilon)(3-2\epsilon)\mu^{3}\overline{\alpha}^{\prime^{3}}}{768(\frac{25\mu^{2}\overline{\alpha}^{\prime^{2}}}{16}+Q_{i}^{2})^{(2.5-\epsilon)}}]
\end{eqnarray}
\begin{eqnarray}
q_{8}=-2AB[\nonumber\frac{1}{(\frac{25\mu^{2}\overline{\alpha}^{\prime^{2}}}{36}+Q_{i}^{2})^{(1-\epsilon)}}-\frac{5(3-2\epsilon)\mu\overline{\alpha}^{\prime}}{6(\frac{25\mu^{2}\overline{\alpha}^{\prime^{2}}}{36}+Q_{i}^{2})^{(1.5-\epsilon)}}&+&\\
\frac{20(4-2\epsilon)(3-2\epsilon)\mu^{2}\overline{\alpha}^{\prime^{2}}}{(27\frac{25\mu^{2}\overline{\alpha}^{\prime^{2}}}{36}+Q_{i}^{2})^{(2-\epsilon)}}-\frac{(5-2\epsilon)(4-2\epsilon)(3-2\epsilon)\mu^{3}\overline{\alpha}^{\prime^{3}}}{27(\frac{25\mu^{2}\overline{\alpha}^{\prime^{2}}}{36}+Q_{i}^{2})^{(2.5-\epsilon)}}]
\end{eqnarray}

\begin{eqnarray}
q_{9}=-2AD^{\prime}[\nonumber\frac{1}{4(\frac{9\mu^{2}\overline{\alpha}^{\prime^{2}}}{16}+Q_{i}^{2})^{(1-\epsilon)}}-\frac{5(3-2\epsilon)\mu\overline{\alpha}^{\prime}}{16(\frac{9\mu^{2}\overline{\alpha}^{\prime^{2}}}{16}+Q_{i}^{2})^{(1.5-\epsilon)}}&+&\\
\nonumber\frac{(4-2\epsilon)(3-2\epsilon)\mu^{2}\overline{\alpha}^{\prime^{2}}}{8(\frac{9\mu^{2}\overline{\alpha}^{\prime^{2}}}{16}+Q_{i}^{2})^{(2-\epsilon)}}-\frac{13(5-2\epsilon)(4-2\epsilon)(3-2\epsilon)\mu^{3}\overline{\alpha}^{\prime^{3}}}{768(\frac{9\mu^{2}\overline{\alpha}^{\prime^{2}}}{16}+Q_{i}^{2})^{(2.5-\epsilon)}}&+&\\\frac{(6-2\epsilon)(5-2\epsilon)(4-2\epsilon)(3-2\epsilon)\mu^{4}\overline{\alpha}^{\prime^{4}}}{1536(\frac{9\mu^{2}\overline{\alpha}^{\prime^{2}}}{16}+Q_{i}^{2})^{(3-\epsilon)}}]
\end{eqnarray}
\begin{eqnarray}
q_{10}=2BD^{\prime}[\nonumber\frac{1}{4(\frac{49\mu^{2}\overline{\alpha}^{\prime^{2}}}{144}+Q_{i}^{2})^{(1-\epsilon)}}-\frac{25(3-2\epsilon)\mu\overline{\alpha}^{\prime}}{48(\frac{49\mu^{2}\overline{\alpha}^{\prime^{2}}}{144}+Q_{i}^{2})^{(1.5-\epsilon)}}&+&\\
\nonumber\frac{151(4-2\epsilon)(3-2\epsilon)\mu^{2}\overline{\alpha}^{\prime^{2}}}{864(\frac{25\mu^{2}\overline{\alpha}^{\prime^{2}}}{144}+Q_{i}^{2})^{(2-\epsilon)}}-\frac{27.66(5-2\epsilon)(4-2\epsilon)(3-2\epsilon)\mu^{3}\overline{\alpha}^{\prime^{3}}}{768(\frac{25\mu^{2}\overline{\alpha}^{\prime^{2}}}{144}+Q_{i}^{2})^{(2.5-\epsilon)}}&+&\\\nonumber\frac{3.66(6-2\epsilon)(5-2\epsilon)(4-2\epsilon)(3-2\epsilon)\mu^{4}\overline{\alpha}^{\prime^{4}}}{1152(25\frac{\mu^{2}\overline{\alpha}^{\prime^{2}}}{144}+Q_{i}^{2})^{(3-\epsilon)}}&-&\\\frac{(7-2\epsilon)(6-2\epsilon)(5-2\epsilon)(4-2\epsilon)(3-2\epsilon)\mu^{4}\overline{\alpha}^{\prime^{4}}}{10368(25\frac{\mu^{2}\overline{\alpha}^{\prime^{2}}}{144}+Q_{i}^{2})^{(3.5-\epsilon)}}]
\end{eqnarray}

\textbf{Expressions for $r_{i}$ s:}

\begin{equation}
r_{1}=3c_{1}^{\prime^{2}}(1+\frac{m_{i}}{m_{j}})^{-2}(4\mu^{2}\overline{\alpha}^{\prime^{2}})^{\epsilon-2}\left(2-2\epsilon\right)
\end{equation}

\begin{eqnarray}
r_{2}=3A^{2}(1+\frac{m_{i}}{m_{j}})^{-2}[\nonumber\left(2-2\epsilon\right)(4\mu^{2}\overline{\alpha}^{\prime^{2}})^{\epsilon-2}-3\mu\overline{\alpha}^{\prime}(3-2\epsilon)^{2}(\mu^{2}\overline{\alpha}^{\prime^{2}})^{\epsilon-2.5}&+&\\ 0.75\mu^{2}\overline{\alpha}^{\prime^{2}}(4-2\epsilon)^{2}(3-2\epsilon)(\mu^{2}\overline{\alpha}^{\prime^{2}})^{\epsilon-3}]
\end{eqnarray}

\begin{eqnarray}
r_{3}=3B^{2}(1+\frac{m_{i}}{m_{j}})^{-2}[\nonumber(\frac{4\mu^{2}\overline{\alpha}^{\prime^{2}}}{9})^{\epsilon-2}\left(2-2\epsilon\right)-4\mu\overline{\alpha}^{\prime}(3-2\epsilon)^{2}(\frac{4\mu^{2}\overline{\alpha}^{\prime^{2}}}{9})^{\epsilon-2.5}&+&\\\nonumber \frac{16\mu^{2}\overline{\alpha}^{\prime^{2}}}{9}(4-2\epsilon)^{2}(3-2\epsilon)(\frac{4\mu^{2}\overline{\alpha}^{\prime^{2}}}{9})^{\epsilon-3}&-&\\\nonumber \frac{8\mu^{3}\overline{\alpha}^{\prime^{3}}}{81}(5-2\epsilon)^{2}(4-2\epsilon)(3-2\epsilon)(\frac{4\mu^{2}\overline{\alpha}^{\prime^{2}}}{9})^{\epsilon-3.5}&+&\\ \frac{4\mu^{4}\overline{\alpha}^{\prime^{4}}}{729}(6-2\epsilon)^{2}(5-2\epsilon)(4-2\epsilon)(3-2\epsilon)(\frac{4\mu^{2}\overline{\alpha}^{\prime^{2}}}{9})^{\epsilon-4}]
\end{eqnarray}

\begin{eqnarray}
r_{4}=3D^{\prime{2}}(1+\frac{m_{i}}{m_{j}})^{-2}[\nonumber(\frac{\mu^{2}\overline{\alpha}^{\prime^{2}}}{4})^{\epsilon-2}\left(2-2\epsilon\right)-\frac{9\mu\overline{\alpha}^{\prime}}{32}(3-2\epsilon)^{2}(\frac{\mu^{2}\overline{\alpha}^{\prime^{2}}}{4})^{\epsilon-2.5}&+&\\\nonumber \frac{57\mu^{2}\overline{\alpha}^{\prime^{2}}}{32}(4-2\epsilon)^{2}(3-2\epsilon)(\frac{\mu^{2}\overline{\alpha}^{\prime^{2}}}{4})^{\epsilon-3}&-&\\\nonumber \frac{19\mu^{3}\overline{\alpha}^{\prime^{3}}}{512}(5-2\epsilon)^{2}(4-2\epsilon)(3-2\epsilon)(\frac{\mu^{2}\overline{\alpha}^{\prime^{2}}}{4})^{\epsilon-3.5}&+&\\\nonumber \frac{21\mu^{4}\overline{\alpha}^{\prime^{4}}}{6144}(6-2\epsilon)^{2}(5-2\epsilon)(4-2\epsilon)(3-2\epsilon)(\frac{\mu^{2}\overline{\alpha}^{\prime^{2}}}{4})^{\epsilon-4}&-&\\\frac{3\mu^{5}\overline{\alpha}^{\prime^{5}}}{12288}(7-2\epsilon)^{2}(6-2\epsilon)(5-2\epsilon)(4-2\epsilon)(3-2\epsilon)(\frac{4\mu^{2}\overline{\alpha}^{\prime^{2}}}{9})^{\epsilon-4.5}]
\end{eqnarray}
\begin{equation}
r_{5}=2c_{1}^{\prime}A(1+\frac{m_{i}}{m_{j}})^{-2}[3(\frac{9\mu^{2}\overline{\alpha}^{\prime^{2}}}{4})^{\epsilon-2}\left(2-2\epsilon\right)-1.5\mu\overline{\alpha}^{\prime}(3-2\epsilon)^{2}(\frac{9\mu^{2}\overline{\alpha}^{\prime^{2}}}{4})^{\epsilon-2.5}]
\end{equation}
\begin{eqnarray}
r_{6}=2c_{1}^{\prime}B(1+\frac{m_{i}}{m_{j}})^{-2}[\nonumber3(\frac{16\mu^{2}\overline{\alpha}^{\prime^{2}}}{9})^{\epsilon-2}\left(2-2\epsilon\right)&-&\\\nonumber2\mu\overline{\alpha}^{\prime}(3-2\epsilon)^{2}(\frac{16\mu^{2}\overline{\alpha}^{\prime^{2}}}{9})^{\epsilon-2.5}&+&\\\frac{2\mu^{2}\overline{\alpha}^{\prime^{2}}}{9}(4-2\epsilon)^{2}(3-2\epsilon)(\frac{16\mu^{2}\overline{\alpha}^{\prime^{2}}}{9})^{\epsilon-3}]
\end{eqnarray}
\begin{eqnarray}
r_{7}=2c_{1}^{\prime}D^{\prime}(1+\frac{m_{i}}{m_{j}})^{-2}[\nonumber(\frac{25\mu^{2}\overline{\alpha}^{\prime^{2}}}{16})^{\epsilon-2}\frac{3\left(2-2\epsilon\right)}{4}&-&\\\nonumber\frac{9\mu\overline{\alpha}^{\prime}(3-2\epsilon)^{2}}{16}(\frac{25\mu^{2}\overline{\alpha}^{\prime^{2}}}{16})^{\epsilon-2.5}&+&\\\nonumber\frac{3\mu^{2}\overline{\alpha}^{\prime^{2}}}{32}(4-2\epsilon)^{2}(3-2\epsilon)(\frac{25\mu^{2}\overline{\alpha}^{\prime^{2}}}{16})^{\epsilon-3}&-&\\\frac{3\mu^{3}\overline{\alpha}^{\prime^{3}}}{768}(5-2\epsilon)^{2}(4-2\epsilon)(3-2\epsilon)(\frac{25\mu^{2}\overline{\alpha}^{\prime^{2}}}{16})^{\epsilon-3.5}]
\end{eqnarray}
\begin{eqnarray}
r_{8}=-2AB(1+\frac{m_{i}}{m_{j}})^{-2}[\nonumber(\frac{25\mu^{2}\overline{\alpha}^{\prime^{2}}}{36})^{\epsilon-2}3\left(2-2\epsilon\right)&-&\\\nonumber2.5\mu\overline{\alpha}^{\prime}(3-2\epsilon)^{2}(\frac{25\mu^{2}\overline{\alpha}^{\prime^{2}}}{36})^{\epsilon-2.5}&+&\\\nonumber\frac{20\mu^{2}\overline{\alpha}^{\prime^{2}}}{9}(4-2\epsilon)^{2}(3-2\epsilon)(\frac{25\mu^{2}\overline{\alpha}^{\prime^{2}}}{36})^{\epsilon-3}&-&\\\frac{\mu^{3}\overline{\alpha}^{\prime^{3}}}{9}(5-2\epsilon)^{2}(4-2\epsilon)(3-2\epsilon)(\frac{25\mu^{2}\overline{\alpha}^{\prime^{2}}}{36})^{\epsilon-3.5}]
\end{eqnarray}
\begin{eqnarray}
r_{9}=-2AD^{\prime}(1+\frac{m_{i}}{m_{j}})^{-2}[\nonumber0.75(\frac{9\mu^{2}\overline{\alpha}^{\prime^{2}}}{16})^{\epsilon-2}\left(2-2\epsilon\right)&-&\\\nonumber\frac{15\mu\overline{\alpha}^{\prime}}{16}(3-2\epsilon)^{2}(\frac{9\mu^{2}\overline{\alpha}^{\prime^{2}}}{16})^{\epsilon-2.5}&+&\\\nonumber\frac{3\mu^{2}\overline{\alpha}^{\prime^{2}}}{8}(4-2\epsilon)^{2}(3-2\epsilon)(\frac{9\mu^{2}\overline{\alpha}^{\prime^{2}}}{16})^{\epsilon-3}&-&\\\nonumber\frac{13\mu^{3}\overline{\alpha}^{\prime^{3}}}{256}(5-2\epsilon)^{2}(4-2\epsilon)(3-2\epsilon)(\frac{9\mu^{2}\overline{\alpha}^{\prime^{2}}}{16})^{\epsilon-3.5}&+&\\\frac{\mu^{4}\overline{\alpha}^{\prime^{4}}}{512}(6-2\epsilon)^{2}(5-2\epsilon)(4-2\epsilon)(3-2\epsilon)(\frac{9\mu^{2}\overline{\alpha}^{\prime^{2}}}{16})^{\epsilon-4}]
\end{eqnarray}
\begin{eqnarray}
r_{10}=2BD^{\prime}(1+\frac{m_{i}}{m_{j}})^{-2}[\nonumber0.75(\frac{49\mu^{2}\overline{\alpha}^{\prime^{2}}}{144})^{\epsilon-2}\left(2-2\epsilon\right)&-&\\\nonumber\frac{17\mu\overline{\alpha}^{\prime}}{16}(3-2\epsilon)^{2}(\frac{49\mu^{2}\overline{\alpha}^{\prime^{2}}}{144})^{\epsilon-2.5}&+&\\\nonumber\frac{151\mu^{2}\overline{\alpha}^{\prime^{2}}}{288}(4-2\epsilon)^{2}(3-2\epsilon)(\frac{49\mu^{2}\overline{\alpha}^{\prime^{2}}}{144})^{\epsilon-3}&-&\\\nonumber\frac{83\mu^{3}\overline{\alpha}^{\prime^{3}}}{768}(5-2\epsilon)^{2}(4-2\epsilon)(3-2\epsilon)(\frac{49\mu^{2}\overline{\alpha}^{\prime^{2}}}{144})^{\epsilon-3.5}&+&\\\nonumber\frac{33\mu^{4}\overline{\alpha}^{\prime^{4}}}{3456}(6-2\epsilon)^{2}(5-2\epsilon)(4-2\epsilon)(3-2\epsilon)(\frac{9\mu^{2}\overline{\alpha}^{\prime^{2}}}{16})^{\epsilon-4}&-&\\\frac{\mu^{5}\overline{\alpha}^{\prime^{5}}}{3456}(7-2\epsilon)^{2}(6-2\epsilon)(5-2\epsilon)(4-2\epsilon)(3-2\epsilon)(\frac{49\mu^{2}\overline{\alpha}^{\prime^{2}}}{144})^{\epsilon-4.5}]
\end{eqnarray}
\textbf{Expressions for $p_{i}$ s:}

\begin{equation}
p_{1}=\frac{c_{1}^{\prime}}{(\frac{\mu^{2}\overline{\alpha}^{\prime^{2}}}{4}+Q_{i}^{2})^{\frac{(3-\epsilon)}{2}}}
\end{equation}
\begin{equation}
p_{2}=A[\frac{1}{(\frac{\mu^{2}\overline{\alpha}^{\prime^{2}}}{4}+Q_{i}^{2})^{\frac{(3-\epsilon)}{2}}}-\frac{(3-2\epsilon)\mu\overline{\alpha}^{\prime}}{2(\frac{\mu^{2}\overline{\alpha}^{\prime^{2}}}{4}+Q_{i}^{2})^{\frac{(4-\epsilon)}{2}}}]
\end{equation}
\begin{eqnarray}
p_{3}=B[\nonumber\frac{1}{(\frac{\mu^{2}\overline{\alpha}^{\prime^{2}}}{9}+Q_{i}^{2})^{\frac{(3-\epsilon)}{2}}}-\nonumber\frac{0.67(3-2\epsilon)\mu\overline{\alpha}^{\prime}}{(\frac{\mu^{2}\overline{\alpha}^{\prime^{2}}}{9}+Q_{i}^{2})^{\frac{(4-\epsilon)}{2}}}&-&\\\frac{2(4-2\epsilon)(3-2\epsilon)\mu^{2}\overline{\alpha}^{\prime^{2}}}{27(\frac{\mu^{2}\overline{\alpha}^{\prime^{2}}}{9}+Q_{i}^{2})^{\frac{(5-\epsilon)}{2}}}]
\end{eqnarray}
\begin{eqnarray}
p_{4}=D^{\prime}[\nonumber\frac{0.25}{(\frac{\mu^{2}\overline{\alpha}^{\prime^{2}}}{16}+Q_{i}^{2})^{\frac{(3-\epsilon)}{2}}}-\nonumber\frac{3(3-2\epsilon)\mu\overline{\alpha}^{\prime}}{16(\frac{\mu^{2}\overline{\alpha}^{\prime^{2}}}{16}+Q_{i}^{2})^{\frac{(4-\epsilon)}{2}}}&-&\\\nonumber\frac{(4-2\epsilon)(3-2\epsilon)\mu^{2}\overline{\alpha}^{\prime^{2}}}{32(\frac{\mu^{2}\overline{\alpha}^{\prime^{2}}}{16}+Q_{i}^{2})^{\frac{(5-\epsilon)}{2}}}&-&\\\frac{(5-2\epsilon)(4-2\epsilon)(3-2\epsilon)\mu^{3}\overline{\alpha}^{\prime^{3}}}{768(\frac{\mu^{2}\overline{\alpha}^{\prime^{2}}}{16}+Q_{i}^{2})^{\frac{(6-\epsilon)}{2}}}]
\end{eqnarray}
\textbf{Expressions for $p_{i}^{\prime}$ s }:
\begin{equation}
p_{1}^{\prime}=n_{1}\times\left(\overline{b}^{\prime}\mu\right)^{\frac{2}{3}}
\end{equation}
\begin{equation}
p_{2}^{\prime}=n_{2}\times\left(\overline{b}^{\prime}\mu\right)^{\frac{2}{3}}
\end{equation}
\begin{equation}
p_{3}^{\prime}=n_{3}\times\left(\overline{b}^{\prime}\mu\right)^{\frac{2}{3}}
\end{equation}
\begin{equation}
p_{4}^{\prime}=n_{4}\times\left(\overline{b}^{\prime}\mu\right)^{\frac{2}{3}}
\end{equation}
\begin{equation}
p_{5}^{\prime}=n_{5}\times\left(\overline{b}^{\prime}\mu\right)^{\frac{2}{3}}
\end{equation}
\begin{equation}
p_{6}^{\prime}=n_{6}\times\left(\overline{b}^{\prime}\mu\right)^{\frac{2}{3}}
\end{equation}
\begin{equation}
p_{7}^{\prime}=n_{7}\times\left(\overline{b}^{\prime}\mu\right)^{\frac{2}{3}}
\end{equation}
\begin{equation}
p_{8}^{\prime}=n_{8}\times\left(\overline{b}^{\prime}\mu\right)^{\frac{2}{3}}
\end{equation}
Each of the constants $n_{1},n_{2},....,n_{8}$  are  different for different  mesons and they have been obtained by numerical integration.
\end{document}